\documentclass[twocolumn,showpacs,superscriptaddress,nobibnotes,floatfix,aps,prb,citeautoscript]{revtex4-1}
\usepackage{rotating}
\usepackage{amsmath}
\usepackage{amsfonts}
\usepackage{amssymb}
\usepackage{mattens}
\usepackage{epsfig}
\usepackage{float}
\usepackage{bm}
\usepackage{color}
\usepackage{graphicx}
\usepackage[active]{srcltx}
\usepackage[sort&compress]{natbib}

\newcommand{\mat}[1]{\Sb*{#1}}  
\newcommand{\qq}{\bm{q}}
\newcommand{\rr}{\bm{r}}
\newcommand{\kk}{\bm{k}}
\newcommand{\KK}{\bm{K}}
\newcommand{\tkk}{\tilde{{\bm k}}}
\newcommand{\tKK}{\tilde{{\bm K}}}
\newcommand{\nkbz}{N_{\kk}^{\text{\tiny BZ}}}
\newcommand{\nkibz}{N_{\kk}^{\text{\tiny IBZ}}}
\newcommand{\GG}{{\bm G}}
\newcommand{\w}{\omega}
\newcommand{\x}{\! \times \!}

\newcommand{\grenoble}{Institut N\'eel, Universit\'e Joseph Fourier
  and CNRS, 38042 Grenoble Cedex 9, France}

\newcommand{\lpmcn}{Universit\'e de Lyon, F-69000 Lyon, France and LPMCN, CNRS UMR 5586, 
  Universit\'e de Lyon 1, F-69622 Villeurbanne Cedex, France}
\newcommand{\palaiseau}{Laboratoire des Solides Irradi\'es and ETSF,
  \'Ecole Polytechnique, CEA-DSM, CNRS, 91128 Palaiseau, France}
\newcommand{\roma}{Istituto di Struttura della Materia (ISM),
  Consiglio Nazionale delle Ricerche, Via Salaria Km 29.5, CP 10,
  00016 Monterotondo Stazione, Italy}

\begin{document}
\title{Speeding up the solution of the Bethe-Salpeter equation by a double-grid method and Wannier interpolation}

\author{David Kammerlander}
\email{kammerlander.david@gmail.com}
\affiliation{\grenoble}
\affiliation{\lpmcn}
\author{Silvana Botti}
\affiliation{\lpmcn}
\affiliation{\palaiseau}
\author{Miguel~A.~L Marques}
\affiliation{\lpmcn}
\author{Andrea Marini}
\affiliation{\roma}
\author{Claudio Attaccalite}
\affiliation{\grenoble}

\begin{abstract}
The Bethe-Salpeter equation is a widely used approach to describe
optical excitations in bulk semiconductors. It leads to spectra that
are in very good agreement with experiment, but the price to pay for
such accuracy is a very high computational burden. One of the main
bottlenecks is the large number of $k$-points required to obtain
converged spectra. In order to circumvent this problem we propose a
strategy to solve the Bethe-Salpeter equation based on a double-grid
technique coupled to a Wannier interpolation of the Kohn-Sham band
structure. This strategy is then benchmarked for a particularly
difficult case, the calculation of the absorption spectrum of
GaAs, and for the well studied case of Si. The considerable gains observed in these cases fully validate our
approach, and open the way for the application of the Bethe-Salpeter
equation to large and complex systems.
\end{abstract}

      
\maketitle


Optical spectra are important for the characterization and prediction
of material properties, as optical excitations are at the core of
e.g., light-emitting devices, laser technology, and photovoltaics. For
extended systems many-body perturbation
theory~\cite{onida:02,botti-review:07}, a Green's function based
approach, is the most accurate method to calculate optical
properties. Perhaps inevitably, it is also one of the most
computationally costly methods available to the community. It involves
the solution of an equation of motion for the two-particle Green's
function, the Bethe-Salpeter equation (BSE), that describes coupled
and correlated electron-hole excitations~\cite{albrecht:98,*benedict:98,*rohlfing:98,*PhysRevLett.33.582}.

The standard numerical techniques used to solve the BSE are based on
an expansion of the relevant quantities in electron-hole states
(needing therefore both filled and empty states), and require a very
dense $k$-point sampling of the Brillouin zone (BZ). Typically, the
number of electron-hole states used in the expansion can be relatively
small if one is only interested in the visible spectra, but the number
of $k$-points can easily reach several thousands. Some approaches have
been put forward to reduce the computational burden of the BSE. For
example, the number of k-points can be reduced by interpolating the
interaction integrals in $k$-space~\cite{rohlfing:00}, while recent
implementations allow for the complete exclusion of empty
states~\cite{rocca:12}.

It is well known that optical spectra are very sensitive to the
$k$-point sampling~\cite{albrecht:99,rocca:12,aguilera:11}. A common
approach to alleviate the problem is the use of arbitrarily shifted $k$-point
grids, that often yield sufficient sampling of the Brillouin zone
while keeping the number of $k$-points manageable. Such a shifted grid, indeed, 
does not use the symmetries of the Brillouin zone and guarantees a maximum 
number of non-equivalent $k$-points accelerating spectrum convergence~\cite{benedict:98b}. 
However, it might induce artificial splitting of normally degenerate states, and thus
produce artifacts in the spectrum~\cite{wirtz:08}, such as the
splitting of some peaks or even the appearance of spurious excitations
in some directions. Of course, these artifacts (slowly) disappear with
increasing density of
$k$-points~\cite{albrecht:99,rocca:12}, and consequent
increase of the computational burden. In view of that a very
dense $k$-point sampling is crucial to obtain an accurate lineshape,
including the correct peak positions~\cite{albrecht:99, hahn:05}, but
very hard to achieve in practical calculations.

In this Article, we propose a new strategy to solve the BSE equation
that alleviates the need for dense $k$-point grids. The
independent-particle part of the BSE is first evaluated on a very
dense $k$-grid ($40\x40\x40$ in the example below) by making use of
Wannier interpolation~\cite{marzari:97}. The BSE is then solved in a
unshifted coarse $k$-grid ($10\x10\x10$ in the example below) using a
double-grid technique to take into account the fast changing
independent-particle contribution. This approach is simple to
implement, and leads to a considerable gain in computational time.

In the following, we start by presenting a short review of the
theoretical ingredients for the description of optical spectra within
the BSE~\cite{rohlfing:00, strinati, *bussi:04}. We then
discuss our approach and prove its usefulness with a notoriously
difficult example, the calculation of the optical absorption
spectrum of the standard semiconductor GaAs. The subsequent discussion
of bulk silicon concludes the benchmark.


The optical absorption spectrum is described by the imaginary part of
the macroscopic dielectric function $\epsilon_{\textrm M}(\w)$ in the
long wavelength limit, which in turn can be obtained from the
two-point contraction of the reducible four-point polarizability $L$,
\begin{equation} \label{eq:eps_macro}
\epsilon_{\textrm M}(\w) = 1 - \lim_{\qq\rightarrow0} v(\qq) \bm{\lambda} \int \! d\rr d\rr' e^{-i\qq(\rr-\rr')} L(\rr,\rr, \rr',\rr';\w), 
\end{equation}
with the Coulomb potential $v=4\pi/(\GG+\qq)^2$, the transferred momentum $\qq$, and $\bm{\lambda}$ the 
direction of light polarization. The
quantity $L$ satisfies the BSE, a Dyson like equation,
\begin{multline} \label{eq:bse_realspace}
  L(1,2,3,4) = L^0(1,2,3,4) + \\
  \int d(5678) L^0(1,2,5,6)\, \Xi(5,6,7,8)\, L(7,8,3,4),
\end{multline}
with the abbreviation of space, spin and time coordinates $(1) =
(\rr_1, \sigma_1, t_1)$, and where $L^0(1,2,3,4)=i G(1,3) G(4,2)$ is
the independent particle polarizability, expressed as a product of
single-particle Green's functions. Equation~\eqref{eq:bse_realspace}
describes the effects of the electron-hole interaction mediated by the
BSE kernel $\Xi=\overline{v} - W$ that is composed of (i) a bare,
repulsive short-range \textit{exchange term}, that includes the
microscopic components of the Coulomb interaction,
i.e. $\overline{v}_{\GG \neq 0}=v_{\GG}; \overline{v}_{\GG=0}=0$, and
(ii) an attractive, static screened Coulomb potential $W$,
the \textit{direct term}, arising from the variation of the
self-energy. 
Dynamical effects due to the self-energy influence both the (single) 
quasiparticle renormalization and the excitonic two-body 
interaction $W$~\cite{bechstedt:97, marini:03}. In response
calculations for semiconductors they partially cancel each other, 
which justifies the commonly 
employed approximation of a static $W$ and neglected quasiparticle 
renormalization, but in general 
this is not true for metals~\cite{marini:03}.

As the interaction is instantaneous, only two time-variables 
of the initial four remain. And, due to the time-translation invariance, 
$L$ and $L^0$ depend only on the relative time-difference. A
time-energy Fourier transformation then turns the polarizability into
a function of a single frequency $L(1,2,3,4;\w)$ where from now on
$(1)=(\rr_1,\sigma_1)$ only.

By taking advantage of the two-particle nature of the BSE, all
expressions are conveniently written on the basis of the electron-hole
vertical transition space composed of $N_v$ valence bands, $N_c$
conduction bands, and $\nkbz$ $k$-points in the whole BZ. The dimension of
this basis is $4 \x N_v \x N_c \x \nkbz$. In the case of vanishing
spin-orbit coupling, the BSE can be separated into the two subspaces of
singlet and triplet excitons~\cite{rohlfing:00}, each of them with
dimension $2 \x N_v \x N_c \x \nkbz$. The basis sets that span these
subspaces are constructed from pairs of single-particle states
$\phi_{n,\kk}$ with $n$ as band index and $\kk$ as $k$-point and spin
variable, such that
\begin{equation}\label{eq:transition_basis}
 \Phi_{\KK}(\rr_1,\rr_2) = \phi_{c,\kk}(\rr_1) \cdot \phi^*_{v,\kk}(\rr_2),
\end{equation}
with the short hand notation $\KK=(c,v,\kk)$, and $c$ and $v$ running
over indices of conduction and valence bands, respectively.

In this basis the polarizability is written as 
\begin{multline}\label{eq:transition_repres}
 L_{\KK_1,\KK_2}(\w) = \int d\rr_1 d\rr_2 d\rr_3 d\rr_4 \\
             \Phi^*_{\KK_1}(\rr_1,\rr_2) L(1,2,3,4;\w) \Phi_{\KK_2}(\rr_3,\rr_4) ,
\end{multline}
and $L^0$, that is now diagonal, reads
\begin{equation}\label{eq:l0_transition_repres}
 L^0_{\KK_1,\KK_2}(\w) = \frac{f_{c_1 \kk_1} - f_{v_1 \kk_1}} {\epsilon_{c_1 \kk_1} - \epsilon_{v_1 \kk_1} - \w -i\eta} \delta_{\KK_1,\KK_2},
\end{equation}
where $f$ denotes the occupation number. The infinitesimal $\eta$
shifts the pole $\w = \epsilon_{c_1 \kk_1} - \epsilon_{v_1 \kk_1}$
away from the real axis, and is thus responsible for a finite
life-time of the excitation.

In this basis, the BSE becomes a matrix equation
\begin{equation} \label{eq:bse}
  L_{\KK_1,\KK_2} = L^0_{\KK_1,\KK_2} +  L^0_{\KK_1,\KK_3} 
                    \Xi_{\KK_3,\KK_4} L_{\KK_4,\KK_2},
\end{equation}
where we used Einstein's notation for summations over repeated
indices in the tensor products, and omitted the explicit energy
dependence for clarity. In the following we adopt the notation of $\mat{O}$ 
for the matrix representation in the electron-hole basis of an arbitrary operator $O$.

It can be shown that the kernel $\mat{\Xi}$ couples pairs of excitations
$(vc)$ with $(v'c')$, but also $(vc)$ with $(c'v')$, leading to the
so-called resonant and coupling terms,
respectively~\cite{rohlfing:00}. Here we make use of the so-called
Tamm-Dancoff approximation, and neglect the latter. We thus arrive at
a Hilbert space of dimension $ N_v \x N_c \x \nkbz$, regardless of the
symmetries of the $k$-grid. Note that these are standard
approximations for the solution of the BSE.

Equation~\eqref{eq:bse} can be solved symbolically, yielding for each
frequency $\w$
\begin{equation} \label{eq:bse_invert}
   L_{\KK_1,\KK_2} =  [ \mat{1 - L^0  \Xi} ]^{-1}_{\KK_1,\KK_3} \, L^0_{\KK_3,\KK_2} .
\end{equation}
To circumvent the inversion of $\mat{1 - L^0 \Xi}$, the usual procedure is
to rewrite the matrix into a two-particle Hamiltonian whose
diagonalization gives the excitonic eigensystem used to express the
polarizability for \textit{all} frequencies at once.

As it will become clear in the following, we stick to the inversion
scheme by taking advantage of the series expansion of
Eq.~\eqref{eq:bse_invert},
\begin{equation} \label{eq:bse_inv_expand} 
  L_{\KK_1,\KK_2}(\w) = \sum_m  [\mat{ L^0(\w) \Xi}]^m _{\KK_1,\KK_3} 
                     \, L^0_{\KK_3,\KK_2}(\w) ,
\end{equation}
that is interrupted at convergence. If, however, a convergence is not attained, 
i.e. the assumption of the expandability of Eq.~\eqref{eq:bse_invert} is 
falsified \textit{a posteriori}, we perform the full inversion. 

The solution of Eq.~\eqref{eq:bse_inv_expand} has two distinct
bottlenecks in terms of computational cost. First, the calculation of
matrix $\mat{\Xi}$ is very time consuming. Second, its storage
needs large quantities of memory. In view of that, reducing the number
of $k$-points is a major issue. To this end, Rohlfing and
Louie~\cite{rohlfing:00} employed a double-grid technique where the
kernel $\mat{\Xi}$ is calculated on a coarse grid with its subsequent
interpolation onto a fine $k$-point mesh where the BSE is solved. This
approach helps reducing the time necessary to compute $\mat{\Xi}$, but it is
less helpful to save memory, since it requires the storage of the
computed and interpolated matrix elements of the kernel. Its use is
justified with the authors' observation that $\mat{\Xi}$ varies little with
respect to the $k$-points, as the single-particle wavefunctions
$\phi_{n\kk}$ are quite robust with respect to $\kk$ (with the
possible exception of sudden band crossings). In a similar spirit it has been 
proved~\cite{marini:01, adragna:03} for the random phase
approximation (RPA), that $\mat{L^0}$ is a rapidly varying quantity
and that could be correctly evaluated by performing additional Monte
Carlo integrations on a large number of random $\kk$ points. Note that
$\mat{L^0}$ can in principle be easily calculated for a large number of
$k$-points and bands.

By taking into account these observations, we define in our approach
two grids: (i)~a coarse one with points $\kk$ on which we calculate
\textit{and} store $\mat{\Xi}$ and solve Eq.~\eqref{eq:bse_inv_expand}, and
(ii)~a fine $k$-grid with vectors $\tkk$ on which we compute
$\mat{L^0}$. The mapping of $L^0_{\tKK_1,\tKK_2 }$ to $L^0_{\KK_1,\KK_2 }$
is performed through a double-grid technique~\cite{ono:99} with a suitably chosen
interpolation for the kernel. To simplify our approach we use the
simplest zeroth order interpolation, that leads to averaging the
finely resolved $\mat{L^0}$ in a neighborhood around each point of the
coarse grid. Consequently, this technique is expected to work if
the oscillator strengths and $\mat{\Xi}$ are smoothly varying functions of $\tkk$.
In practice, we define:
\begin{equation} \label{eq:l0_mean}
 L^0(\w)_{\KK_1,\KK_2 } = \frac{1}{N_{\tkk}} \sum_{\tkk \in \mathcal{D}_{\kk}} 
  \frac{f_{c_1 \tkk} - f_{v_1 \tkk}} {\epsilon_{c_1 \tkk} - \epsilon_{v_1 \tkk} - \w -i\eta} \delta_{\KK_1,\KK_2},
\end{equation}
where $N_{\tkk}$ is the number of $k$-points of the fine grid in the
domain $\mathcal{D}_{\kk}$ around $\kk$ of the coarse grid. We would
like to note that we are averaging the polarization $\mat{L^0}$ that has
poles at the particle excitation energies, which is not equivalent to
averaging the particle excitation energies that appear in the diagonal of
the excitonic Hamiltonian. An arbitrary $k$-point resolution of $\mat{L^0}$
is possible once the respective single-particle energies
$\epsilon_{n\tkk}$ are available.

In general the calculation of quasiparticle states on the $\tkk$ grid
are not practical.
Fortunately, there is a solution to this problem that
relies on the interpolation of the (quasiparticle) electronic
structure to a dense $k$-grid using maximally localized Wannier
functions~\cite{marzari:97}. In this method, $\epsilon_{n\tkk}$ at an
arbitrary $k$-point $\tkk$ is the result of (i) a rotation of the
initial quasiparticle Hamiltonian into the Wannier basis, (ii) its
Fourier interpolation to the fine grid of $k$-points, and (iii) the
diagonalization of the resulting Hamiltonian~\cite{hamann:09}. Note
that, even if this procedure leads to an expression that has the form
of a Slater-Koster tight-binding interpolation~\cite{slater:54}, the
obtained single-particle energies are calculated directly from the
underlying \textit{ab initio} eigensystem rather than just fitted to
it.

\begin{figure}
 \begin{center}
  \includegraphics[width=0.99\columnwidth,clip]{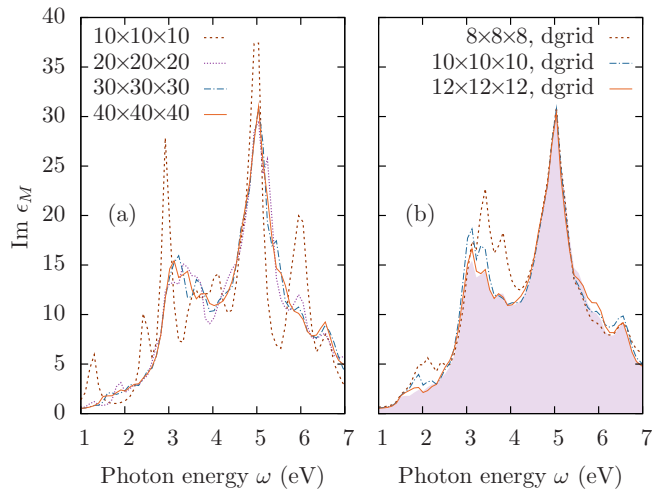} 
  \caption{\label{fig:gaas_spectrum_rpa} (Color online) Calculated RPA
    absorption spectra starting from $GW$ corrected bands for bulk
    GaAs with $N_v=N_c=2$ for various $k$-grids, without (a) and with (b) double-grid
    method. The shaded area in (b) shows the
    spectrum corresponding to the $40\x40\x40$ $k$-grid of panel (a) for better comparison
    between the double-grid method and the standard calculations.}
 \end{center}
\end{figure}
\begin{figure}
 \begin{center}
  \includegraphics[width=0.99\columnwidth,clip]{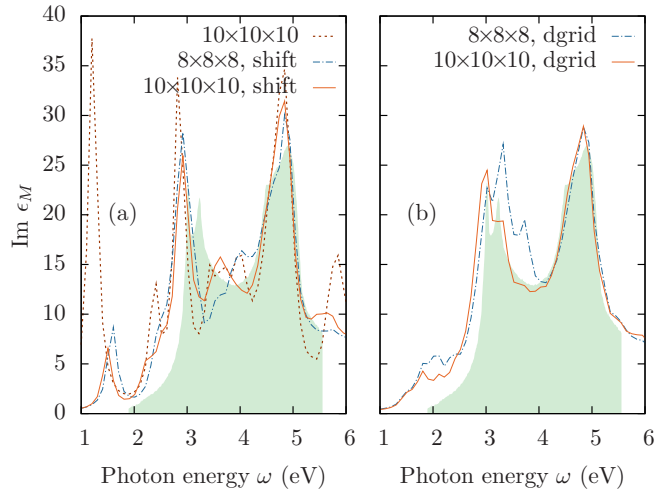} 
  \caption{\label{fig:gaas_spectrum_bse} (Color online) Calculated BSE
    absorption spectra for bulk GaAs with $N_v=2$ and $N_c=3$ using
    several $k$-grids without (a) and with (b) double-grid method. 
    In (a) one spectrum is calculated on a symmetric $10\x10\x10$
    $k$-grid, while the others are on shifted grids to accelerate
    convergence. The shaded area in both panels indicate the
    experimental spectrum at $22$~K~\cite{lautenschlager:87}. }
 \end{center}
\end{figure}
To illustrate the efficiency of our scheme we calculated the optical
spectra of semiconducting GaAs, known for its slow convergence with
respect to the $k$-point sampling~\cite{rohlfing:00, hahn:05,
  benedict:98}. The Kohn-Sham band structure and wavefunctions were
obtained with density functional theory (DFT) within the local density
approximation using norm conserving pseudopotentials with an energy
cutoff of $14$~Hartree and the experimental lattice constant of
$10.68$~Bohr~\cite{cohen:88}.

For the DFT part we utilized the code
\texttt{ABINIT}~\cite{abinit:09}. To obtain the energy bands [used in
  Eq.~\eqref{eq:l0_mean}] we employ the code
\texttt{wannier90}~\cite{wannier90} to perform a Wannier interpolation
to a $40 \x 40 \x 40$ regular $k$-point grid in the whole BZ. Finally,
optical spectra were calculated using the code
\texttt{Yambo}~\cite{yambo} that uses the DFT Kohn-Sham wavefunctions
and the interpolated single-particle energies as input.

It is well known~\cite{godby:87} that
self-energy corrections in GaAs can be simulated by a rigid shift of
the conduction bands. We therefore applied a scissor operator of
$0.9$~eV, that yields an overall agreement of the band dispersions
within 0.1\,eV with the $GW$ corrected bands, and a close agreement with
experimental data as
well~\cite{bimberg:72,aspnes:76,chiang:80,wolford:85,aspnes:86}.

For all spectra in Fig.~\ref{fig:gaas_spectrum_rpa} and
\ref{fig:gaas_spectrum_bse} we included the two highest valence bands and
the two (three for BSE) lowest conduction bands, considering only the resonant part
of the BSE kernel $\mat{\Xi}$, and used a Lorentzian
broadening of $0.1$~eV. Furthermore, we neglected spin-orbit coupling. 
Omitting local field effects (LFE), the non-interacting RPA spectra in 
Fig.~\ref{fig:gaas_spectrum_rpa}(a) are obtained on symmetric Monkhorst-Pack 
(MP) grids~\cite{monkhorst:76}. With increasing 
$k$-point resolution the spectrum converges to two main peaks
at $3.3$~eV and $5.3$~eV. By using our double-grid method, shown in panel
(b), a $12\x12\x12$ symmetric grid yields an
equally well converged spectrum. We observe an excellent agreement
between the latter and the RPA done on a $40\x40\x40$ grid (indicated
by the shaded area).

In general, if one calculates independent-electron transitions
starting from $GW$ corrected bands the oscillator strength of the
absorption spectrum is moved too high in energy compared to the
experiment. The attractive net electron-hole interaction decreases the
energy of the excited states and transfers oscillator strength to
lower energies. This can be seen by comparing the non-interacting and
interacting results of Figs.~\ref{fig:gaas_spectrum_rpa} and
\ref{fig:gaas_spectrum_bse}, respectively.

Figure~\ref{fig:gaas_spectrum_bse}(a) illustrates that a symmetric
$10\x10\x10$ grid alone does not provide enough independent sampling
points for a converged BSE spectrum. Shifting this grid in a direction
different from the high symmetry directions provides 1~000 instead of
only 47 nonequivalent sampling points (see
Table~\ref{tab:kpoints}). This leads to a spectrum that is
sufficiently compatible with experimental results of
Ref.~\onlinecite{lautenschlager:87}. Nevertheless, the low-energy
region (peak at $1.9$~eV) and the region between the two main
transitions at $3.2$~eV and $5.1$~eV are still expected to change on a
denser $k$-grid.  With our double-grid technique the BSE spectrum
is converged even on a symmetric grid of $10\x10\x10$ [see
Fig.~\ref{fig:gaas_spectrum_bse}(b)]. Contrary to the spectrum on an
equally dense, but shifted grid, our spectrum is smooth in between the
main transitions at $3.2$~eV and $5.1$~eV.  It is noteworthy to
mention that, for our scheme, a shifted, coarse $k$-grid does not
improve the convergence of the spectrum of GaAs.

\begin{figure}
 \begin{center}
  \includegraphics[width=0.99\columnwidth,clip]{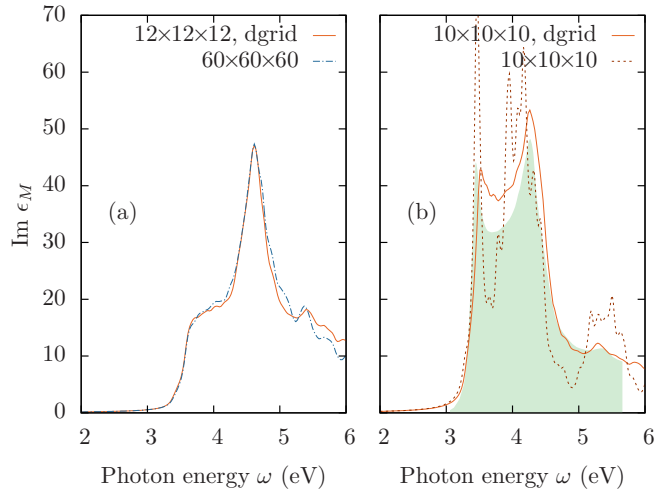} 
  \caption{\label{fig:si_spectra} (Color online) Calculated RPA (a) and BSE (b)
    absorption spectra with LFE starting from $GW$ corrected bands for bulk
    Si with and without double-grid method employing a broadening of $0.05$~eV. 
    For (a) $N_v=N_c=2$ were used, while for (b) $N_v=2$ and $N_c=3$. With the 
    double-grid method both spectra (solid lines) converge faster. 
    The shaded area in (b) shows the experimental spectrum~\cite{lautenschlager:87si}. 
    }
 \end{center}
\end{figure}
In the same fashion we calculated the RPA and BSE absorption spectra of Si, 
shown in Fig.~\ref{fig:si_spectra}. 
To converge the ground state KS energies and wavefunctions with DFT we used 
an energy cutoff of $15$~Ha and a lattice constant of $10.2$~Bohr obtained by crystal 
relaxation~\cite{dalcorso:94}. Similarly to GaAs, we took advantage of the good approximation
of self-energy corrections by a rigid shift of the conduction bands of $0.8$~eV~\cite{rocca:12}. 
For the spectra we employed the two highest valence bands and the two (three for BSE) lowest conduction bands, 
and we included LFE by a dielectric matrix of size $51\x51$. Furthermore, we diminished the Lorentzian 
broadening to $0.05$~eV in order to better resolve the first peak of the BSE spectrum at $3.5$~eV. 

Consequently a high $k$-point resolution of $60\x60\x60$ was necessary to converge 
the RPA spectrum on a symmetric grid without the double-grid method. 
Figure~\ref{fig:si_spectra}(a) illustrates the advantage of the double-grid 
technique as it warrants a converged RPA spectrum on a $12\x12\x12$ grid. 
With this method also the BSE spectrum in Fig.~\ref{fig:si_spectra}(b) converged 
fast on a $10\x10\x10$ grid, while the standard method of diagonalizing the BSE 
Hamiltonian is still far off convergence on the same $k$-grid. Additionally, we 
observe a good agreement of the converged BSE spectrum with experiment~\cite{lautenschlager:87si}
(indicated by the shaded area).

A final remark for Si on the sampling of the dense $k$-points $\tkk$ in 
Eq.~\eqref{eq:l0_mean} is in order now. Instead of using a $40\x40\x40$ regular $k$-grid 
of the full BZ (as for GaAs), we found it favorable to resort to a set of 
four $40\x40\x40$ shifted MP grids that respects the face-centered cubic symmetry of the Si 
crystal. Note that the calculations of the spectra using the double-grid technique were then 
again performed on unshifted, symmetric MP grids.

Although a MP grid that respects the symmetries of the BZ
does not reduce the dimension of the BSE kernel, its use is still
advantageous for three reasons. Firstly, there is no artificial
splitting of degenerate states~\cite{wirtz:08}. Secondly, no
artificial crystal anisotropy is introduced that has to be compensated
by averaging the computed spectra over the three spatial directions of
light polarization~\cite{sottile:03, rocca:12}. Finally, in the
calculation of the exchange term of $\mat{\Xi}$, symmetries of the BZ can be
exploited, which translates in a strong reduction of computational
time. In Table~\ref{tab:kpoints} we have summarized the number of
$k$-points with and without considering symmetries of the BZ.

\begin{table}
\caption{\label{tab:kpoints} Number of $k$-points in the (ir)reducible
  BZ $N_{\kk}^{\text{\tiny (I)BZ}}$ that are used for the calculation
  of the spectra in Figs.~\ref{fig:gaas_spectrum_rpa} -
  \ref{fig:si_spectra}. }
\begin{tabular}{llcrr}
\hline        
\hline       
   calc.      & Figs. & $k$-point grid  &  $\nkibz$   & $\nkbz$ \\
\hline 
   RPA        & \ref{fig:si_spectra}(a)                                    & 60x60x60         &5~216      & 216~000  \\
   RPA        & \ref{fig:gaas_spectrum_rpa}(a)                             & 40x40x40         &1~661      & 64~000  \\
   RPA, dgrid & \ref{fig:gaas_spectrum_rpa}(b), \ref{fig:si_spectra}(a)     & 12x12x12         & 72       & 1~728    \\
   BSE        & \ref{fig:gaas_spectrum_bse}(a)                             & 10x10x10 shifted & 1~000     & 1~000   \\
   BSE, dgrid & \ref{fig:gaas_spectrum_bse}(b), \ref{fig:si_spectra}(b)     & 10x10x10         & 47       & 1~000   \\
\hline        
\end{tabular}
\end{table}

In conclusion, we presented a double-grid method to solve the BSE on a
coarse $k$-point grid, where the average of the strongly varying, but
easily obtainable, independent-particle polarization is used. Converged
spectra are reached for relatively small symmetric $k$-point grids.
This allows for a considerably faster calculation of the BSE
kernel. The single-particle energy bands in a dense $k$-point grid,
the basic ingredient of our method, are not calculated directly, but
are obtained through Wannier interpolation of the electronic
band-structure. As examples, we discussed the convergence of the
absorption spectra of GaAs and Si with respect of the number of $k$-points.
The speed-up is considerable, and opens the way for the solution of
the BSE equation in large, complex systems.

D.~K. and C.~A. are financially supported by the Joseph Fourier university funding program for research (p\^ole Smingue). D.~K. and M.~A.~L.~M. acknowledge financial support from the French ANR (ANR-08-CEXC8-008-01). Computational resources were provided by GENCI (project x2011096017).

\addcontentsline{toc}{chapter}{Bibliography}
\bibliographystyle{apsrev4-1}
%

\end{document}